\documentclass[review,authoryear,12pt]{elsarticle}

\usepackage{epsfig}
\usepackage{lineno}\linenumbers*[1]

\usepackage{amssymb}

\usepackage[ps2pdf,%
a4paper=true,%
breaklinks=true,%
colorlinks=true,%
pdfauthor={First Author et al.},%
pdftitle={Template for manuscripts in Advances in Space Research}%
]{hyperref}

\journal{Advances in Space Research}

\begin{document}

\begin{frontmatter}

\title{Changes in Sea-Level Pressure over South Korea Associated with High-Speed Solar Wind
Events}


\author[kasi,ust]{Il-Hyun Cho}
\author[kasi]{Young-Sil Kwak}
\author[kasi]{Katsuhide Marubashi}
\author[kasi]{Yeon-Han Kim}
\author[kasi]{Young-Deuk Park}
\author[knu]{Heon-Young Chang\corref{cor1}}
\ead{hyc@knu.ac.kr}
\cortext[cor1]{Corresponding author}

\address[kasi]{Korea Astronomy and Space Science Institute, Daejeon, 305-348, Korea}
\address[ust]{University of Science and Technology, Daejeon, 305-348, Korea}
\address[knu]{Dept. of Astronomy and Atmospheric Sciences, Kyungpook National University, Daegu 702-701, Korea}

\begin{abstract}
We explore a possibility that the daily sea-level pressure (SLP) over South Korea responds to the high-speed solar wind event. This is of interest in two aspects: First, if there is a statistical association this can be another piece of evidence showing that various meteorological observables indeed respond to variations in the interplanetary environment. Second, this can be a very crucial observational constraint since most models proposed so far are expected to preferentially work in higher latitude regions than the low latitude region studied here. We have examined daily solar wind speed ${\rm V}$, daily SLP difference ${\rm \Delta SLP}$, and daily ${\rm \log(BV^{2})}$ using the superposed epoch analysis in which the key date is set such that the daily solar wind speed exceeds 800 ${\rm kms^{-1}}$. We find that the daily ${\rm \Delta SLP}$ averaged out of 12 events reaches its peak at day $+1$ and gradually decreases back to its normal level. The amount of positive deviation of ${\rm \Delta SLP}$ is $+2.5$ hPa. The duration of deviation is a few days. We also find that ${\rm \Delta SLP}$ is well correlated with both the speed of solar wind and ${\rm \log(BV^{2})}$. The obtained linear correlation coefficients and chance probabilities with one-day lag for two cases are $r \simeq 0.81$ with $P> 99.9\%$, and $r \simeq 0.84$ with $P> 99.9\%$, respectively. We conclude by briefly discussing future direction to pursue.
\end{abstract}


\end{frontmatter}

\parindent=0.5 cm


\section{Introduction}

Various aspects of solar variability are known to be linked to changes in the Earth's weather and climate on the day-to-day timescale to several tens of year timescale [Cho and Chang 2008; Kniveton et al. 2008; Scafetta and West 2006; Krivova and Solanki 2004; Egorova, Vovk, and Troshichev 2000; Marsh and Svensmark 2000; Svensmark and Friis-Christensen 1997; Pudovkin and Veretenenko 1996a; Tinsley and Heelis 1993; Friis-Christensen and Lassen 1991]. For example, global-average thermospheric total mass density is highly sensitive to solar EUV irradiance, and to the high-latitude electric fields and currents generated by the interaction of the solar wind and the embedded interplanetary magnetic field with the Earth's magnetosphere as well [Weimer et al. 2011; Emmert and Picone 2010; Liu et al. 2010; Kwak et al. 2009; Bruinsma, Tamagnan, and Biancale 2004; Bruinsma et al. 2006; Forbes et al. 2005; Liu and Luhr 2005; Sutton, Forbes, and Nerem 2005]. Solar energetic particles are also known to deplete ozone and to cause other chemical changes in the upper stratosphere and mesosphere [e.g., Reid, Solomon, and Garcia 1991; Crutzen, Isaksen, and Reid 1975]. Moreover, changes in the temperature and dynamics in the troposphere is suggested to correlate with the Earth's magenetic/electric changes corresponding to solar activities [Burns et al. 2007, 2008; Tinsley 2000, 2008; Troshichev 2008; Huth et al. 2007; Baranyi and Ludmany 2005; Veretenenko and Thejll 2004, 2005; Roldugin and Tinsley 2004; Kodera 2003; Boberg and Lundstedt 2002; Yu 2002; Todd and Kniveton 2001; Rycroft, Israelsson, and Price 2000; Egorova, Vovk, and Troshichev 2000; Gabis et al. 2000; van Loon and Shea 1999; van Loon and Labitzke 1988; Pudovkin and Veretenenko 1995; Stozhkov et al. 1995; Tinsley and Heelis 1993; Tinsley and Deen 1991; Venne and Dartt 1990; Page 1989; Tinsley, Brown, and Scherrer 1989; Brown and John 1979; Hoyt and Schatten 1977; Larsen and Kelly 1977; Schuurmans 1965;  Mansurov et al. 1974; Wilcox et al. 1973].

One possible explanation for this link between solar variability and changes in the Earth's weather is that changes in cloud microphysics are caused by variations in the current that flows downward from the ionosphere to land or ocean surface. Observations consistent with this involve changes in surface pressure in the polar regions associated with changes in the ${\rm B_{y}}$ component of the interplanetary magnetic field (IMF), or more precisely changes in the product of ${\rm B_{y}}$ with the solar wind speed, so called the Mansurov effect [Mansurov et al. 1974; Page 1989]. This product causes changes in the polar ionospheric potential, causing changes in the ionosphere-earth current, which affects the production of space charge in layer clouds, with the charges being transferred to droplets and aerosol particles. Variations in the current affect the production of space charge in layer clouds, with the charges being transferred to droplets and aerosol particles. Thus, the changes in electric properties of the atmosphere influence weather and climate. The pressure changes, ${\rm \Delta P}$, are of amplitude a few hPa, and are opposite in the Arctic as compared with the Antarctic. An analysis for the new data set by Burns et al. [2007, 2008] was made with respect to the IMF ${\rm B_{y}}$ component, and demonstrated how the solar wind can modulate the currents in the global electric circuit in the ionosphere and how this modulation can cause changes in tropospheric dynamics, as Tinsley [2000] suggested. There are also many studies that the surface pressure field in high latitude regions shows a variation responding to the geomagnetic storm which may be caused by the variation in the IMF condition such as its intensity and flow speed [Manohar and Subramanian 2008; Bochn\'{i}\v{c}ek et al. 1999; Smirov \& Kononovich 1996; Mustel et al. 1977].

We note that most of reported observational evidence for changes in the lower atmosphere associated with solar activity phenomena is found in  the high magnetic latitude sites, such as Vostok (78.5$^\circ$S, 107$^\circ$E), Sodankyla (67.2$^\circ$N, 26$^\circ$E) [e.g., Burns et al. 2007, 2008; Pudovkin et al. 1996b, 1997; Pudovkin and Baabushkina 1992]. In this short contribution the tropospheric responses to a high-speed solar wind event and related events in the form of sea-level pressure variations at rather low latitude are studied. We investigate whether the sea-level pressure (SLP) over South Korea ($\sim$ 36$^\circ$N, $\sim$ 128$^\circ$E) responds to the high-speed solar wind event consistently as seen in high latitude regions, applying the superposed epoch analysis technique in a statistical treatment. We believe this is an interesting issue for two reasons. First, if there is a statistically significant association this can be another piece of evidence showing that various meteorological observables indeed respond to variations in the interplanetary environment. Second, probably more importantly, this can be a very crucial observational constraint in the sense that most models proposed so far are expected to preferentially work in higher latitude regions than the low latitude region studied here.

We briefly describe data sets in Section 2, and present obtained results in Section 3. We discuss and conclude in Section 4.

\section{Data}

Daily solar wind data is taken from the National Space Science Data Center (NSSDC) OMNIWeb database\footnote{${\rm http://omniweb.gsfc.nasa.gov/}$}, for the time interval from 1986 to present, where the solar wind data have been compiled since 1963 using observed data from 7 satellites including ACE, WIND and IMP. From the time series data of daily solar wind speed we have selected time intervals of twenty-one days whose daily solar wind speed at maximum exceeds 800 ${\rm kms^{-1}}$. The occurrence probability that the daily solar wind speed exceeds 800 ${\rm kms^{-1}}$ is very low, that is, less than 0.1 \%. This event is sometimes called a high-speed solar wind stream (HSS), and these originate from solar coronal holes. The events are further chosen such that recurrent maxima exceeding 800 ${\rm kms^{-1}}$ are separated by at least 31 days to avoid overlapping events. As a result, we end up with 12 high-speed solar wind events. Finally, for the superposed epoch analysis we set the key date (i.e., day number zero) when the daily solar wind speed exceeds 800 ${\rm kms^{-1}}$, as listed in Table 1.

We have used the daily SLP collected from a number of meteorological observation stations distributed over the Republic of Korea during the period from 1986 to present. Korea Meteorological Administration (KMA)\footnote{${\rm http://web.kma.go.kr/eng/index.jsp}$} has observed and tabulated daily surface pressure from a network of 76 ground-based  stations to produce a data set of daily observations. The recorded pressure is further corrected to the one at the sea level by a standard procedure to take the altitude of each station into account. The daily SLP used in the present analysis is given by the spatial average over the 63 stations, whose locations are shown by filled circles in Figure 1, in order to guarantee the temporal homogeneity. Day-to-day variations of SLP recorded at different stations behave in quite a similar pattern. A typical standard deviation resulting from the spatial average for a given day is $ \sim 1$ hPa. Of 12 events only two events are influenced by typhoons, as shown in Table 1. Periods given in Table 1 are based on the official announcement of KMA that concerned typhoons begins/ends to seriously affect meteorological environments of the Korean peninsula.

\section{Results}

In Figure 2, we compare the mean profiles of the solar wind speed and the SLP difference (${\rm \Delta SLP}$) using 12 events to see whether there is a noticeable response of SLP to the high-speed solar wind event. The error bars in both panels denote by the standard error of the mean. In the upper panel the mean profile of the daily solar wind speed is seen to rise rapidly during days from $-2$ to $0$, to reach to the maximum at day 0, and to gradually decreases after the key date back to normal level. The characteristic duration of events is a few days. In the lower panel, we show the mean profiles of ${\rm \Delta SLP}$ averaged by 12 ${\rm \Delta SLP}$, each of which is defined as the difference between the average value over the period of day $-5$ to $-1$ and the daily value. Note this definition is different from that commonly used by, such as, Burns et al. (2007, 2008), Troshichev (2008). One advantage of the definition adopted in our study is that the curve is less deformed by smoothing data, since defining a variation value as the difference between the daily value and the average of some days either side basically involves a moving average operation. One may easily see the response of SLP to the high-speed solar wind event from the bottom panel in the sense that ${\rm \Delta SLP}$ reaches its peak at day $+1$ and gradually decreases back to its normal level. The amount of positive deviation of ${\rm \Delta SLP}$ is $+2.5$ hPa, which is significantly larger than the statistical random fluctuation, $\sim 1$ hPa, even when including two typhoon-contaminated events. The duration of deviation is a few days. We have also carried out the Student's t-test to disprove the null-hypothesis that ${\rm \Delta SLP}$ does not respond to the high-speed solar wind event (in other words, ${\rm \Delta SLP}$ at day $+1$ does not significantly differ from the mean value of the interval from day $-5$ to $-1$). Its resulting false-alarm level is lower than 0.1\%, which allows us to conclude the increase of $+2.5$ hPa in ${\rm \Delta SLP}$ is statistically significant.

In Figure 3, we also compare the mean profile of (${\rm \Delta SLP}$) and that of ${\rm \log(BV^{2})}$ where ${\rm B}$ is the IMF magnetic field intensity, which is taken from the National Space Science Data Center (NSSDC) OMNIWeb database\footnote{${\rm http://omniweb.gsfc.nasa.gov/}$}. Although, the solar wind speed is by itself a good proxy for the geomagnetic disturbance, actual energy input to the Earth's atmosphere from IMF may be directly related to the IMF flow energy density, ${\rm \log(BV^{2})}$ [Lei et al. 2008]. This quantity also includes an information not only on the flow speed carrying the IMF but also on its various aspects of atmospheric input which is mainly due to ${\rm B_{y}}$ and ${\rm B_{z}}$ effects. The error bars in both panels denote by the standard error of the mean. Once again it can be seen that SLP responds in the same way as in the case of solar wind speed.

In Figure 4, we show scatter plots of mean ${\rm \Delta SLP}$ versus the mean speed of solar wind and mean ${\rm \log(BV^{2})}$  in the left and right panel, respectively, which are taken from Figures 2 and 3. Note that in order to take into account the 1-day lag the value of ${\rm \Delta SLP}$ is taken from the following day after other physical quantities are read. A linear regression is fitted to these 20 points. We have found that ${\rm \Delta SLP}$ is well correlated with both the speed of solar wind and ${\rm \log(BV^{2})}$. We calculate Pearson's  linear correlation coefficient $r$ and the chance probability that $r$ has an equal or larger value than its observed in the null hypothesis. The obtained correlation coefficients and chance probabilities with one-day lag for two cases are $r \simeq 0.81$ with $P> 99.9\%,$, and $r \simeq 0.84$ with $P>99.9\%$, respectively. This estimate of significance assumes the data points are independent, which may not be true since many geophysical data sets are self-correlated over extended time intervals.

\section{Discussion and Conclusion}

We perform the superposed epoch analysis to explore a possible response of the SLP over South Korea to the high-speed solar wind event. The average profile of superposed ${\rm \Delta SLP}$ shows a rapid increase up to 2.5 hPa at day $+1$ and a gradual decrease to its normal level, whose key date is defined such that whose daily solar wind speed at maximum exceeds 800 ${\rm kms^{-1}}$. We find that the SLP in a low latitude region shows a measurable response to an encounter of  the high-speed solar wind as seen in the high latitude region.

Most of high-speed solar wind events in this study were produced by flare associated CMEs. The fact that FD events, which are characterized by a decrease of cosmic ray influx, occurred around 9 key dates implies that they are produced by fast CMEs rather than recurrent coronal holes. Therefore, key dates represent not only its high speed but also its strong IMF condition. During $\pm$1 day from the key date, the shock front of the magnetic cloud and the magnetic cloud itself represented by slowly varying field intensity produce Forbush decrease (FD) event. The geomagnetic environment may remain disturbed by the magnetic cloud passing the earth. As mentioned above, both the cosmic ray decrease and the disturbed condition of geomagnetic field are widely accepted as sources of variations in the atmospheric electric current, and their effects has been detected at the troposphere in the form of the surface pressure in high latitude regions. According to our results, FD and/or disturbed geomagnetic condition have likely influence on the tropospheric condition in the low latitude region as well. In Table 1, for comparison we also list dates of FD, which are close to that of our key dates, we found in the Neutron Monitor Database (NMDB)\footnote{${\rm http://http://www.nmdb.eu/nest/search.php}$}.

Further investigation is needed to quantify the possibility that the physical mechanism is a response to the Forbush decreases. For this analysis a large set of FD data may be used and subsampled according to the characteristics. Another direction to pursue is that various meteorological quantities, such as, the temperature or wind speed should be studied, which are uniformly surveyed from a broader and more meteorologically stable area than the Korean peninsula.

\section*{Acknowledgments}
We thank the anonymous referees for careful reading of the manuscript and critical comments which improve the original version of the manuscript. This work was supported by the 'Development of Korean Space Weather Center' of KASI and KASI basic research funds. HYC was supported by the National Research Foundation of Korea Grant funded by the Korean Government(NRF-2010-013-C00017).


\newpage

\begin{figure}
\centering
\noindent\includegraphics[width=30pc]{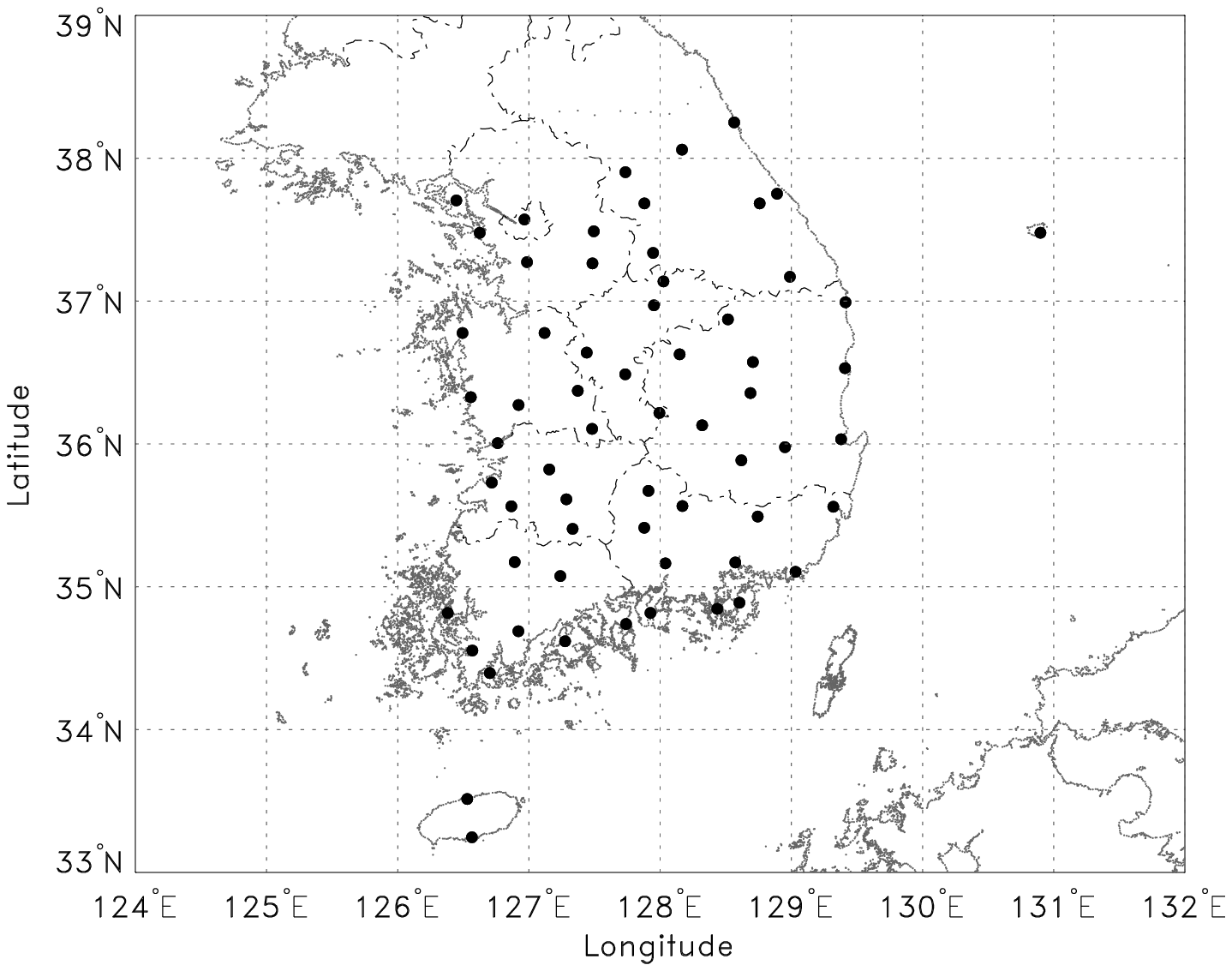} 
\caption{Meteorological stations over the South Korea, where the daily SLP are collected.  The abscissa and ordinate represent longitude ($^\circ E$) and latitude ($^\circ N$), respectively.}
\end{figure}

\begin{figure}
\centering
\noindent\includegraphics[width=30pc]{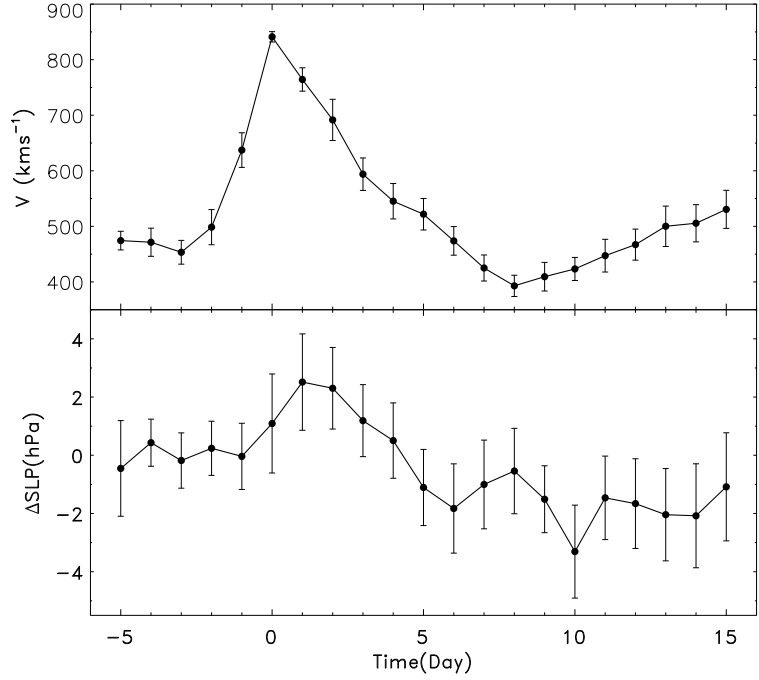} 
\caption{Mean profiles of the mean speed of solar wind (top) and the SLP difference (${\rm \Delta SLP}$) (bottom). We set the day number zero when the daily solar wind speed exceeds 800 ${\rm kms^{-1}}$. In the upper panel the mean profile of the daily solar wind speed is shown to rise rapidly during days from $-2$ to $0$, and gradually decreases after the key date. In the lower panel, we show the mean ${\rm \Delta SLP}$ averaged by 12 ${\rm \Delta SLP}$, each of which is defined as the difference between the average value from day $-5$ to $-1$ and the daily value.}
\end{figure}

\begin{figure}
\centering
\noindent\includegraphics[width=30pc]{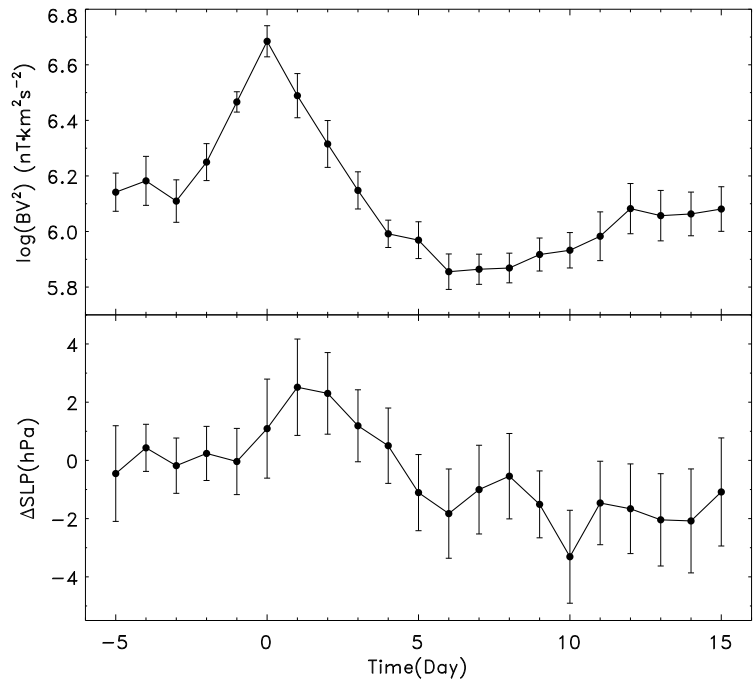} 
\caption{Similar to Figure 2, except that the upper panel is due to ${\rm \log(BV^{2})}$, where ${\rm B}$ is the IMF magnetic field intensity which is directly related to the actual energy input.}
\end{figure}

\begin{figure}
\centering
\noindent\includegraphics[width=24pc, angle=90]{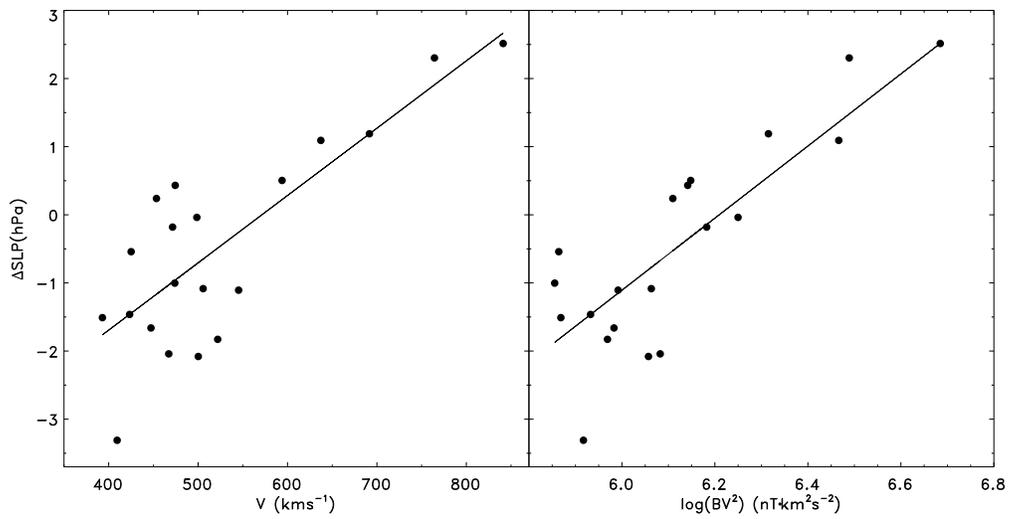} 
\caption{Mean ${\rm \Delta SLP}$ versus the mean speed of solar wind (left) and mean ${\rm \log(BV^{2})}$ (right). Note that the value of ${\rm \Delta SLP}$ is taken from the following day to take into account the 1-day lag. A linear regression is fitted to these 20 points and shown with the solid line.}
\end{figure}

\begin{table}
\begin{center}
\caption{Dates of the high-speed solar wind event (HSE) and Forbush Decrease (FD) close to that, and periods of typhoons passing the Korean peninsula around dates of the high-speed solar wind event according to the official announcement of KMA.}
\begin{tabular}{cccc} \hline
HSE & FD & Typhoon \\ \hline
'86. 2. 8 &           &       &\\ \hline
'89. 3.14 & '89. 3.13 &       &\\ \hline
'89.10.21 & '89.10.20 &       &\\ \hline
'90. 6.12 & '91. 6. 9 &       &\\
          & '91. 6.12 &       &\\ \hline
'92. 5.10 & '91. 5. 9 &       &\\ \hline
'00. 7.15 & '00. 7.15 &       &\\
          & '00. 7.13 &       &\\ \hline
'00.11.11 &           &       &\\ \hline
'03.10.29 & '03.10.29 &       &\\ \hline
'03.12.11 &           &       &\\ \hline
'04. 7.27 & '04. 7.26 & '04. 7.25$\sim$ 8. 1 &\\ \hline
'05. 1.19 & '05. 1.18 &        &\\
          & '05. 1.22 &        &\\ \hline
'05. 9.11 & '05. 9.11 & '05. 8.29$\sim$ 9. 7 &\\ \hline
\end{tabular}
\end{center}
\end{table}

\end{document}